\documentclass[twocolumn,amsmath,amssymb,floatfix]{revtex4}
\usepackage{graphicx}
\usepackage{bm}
\usepackage[usenames]{color}

\newcommand{\beq}{\begin{eqnarray}}
\newcommand{\eeq}{\end{eqnarray}}

\def \be{\begin{equation}}
\def \ee{\end{equation}}
\def \ba{\begin{array}}
\def \ea{\end{array}}
\def \bea{\begin{eqnarray}}
\def \eea{\end{eqnarray}}

\begin{document}

\title{
Weber blockade theory of magnetoresistance oscillations in superconducting strips
}

\author{David Pekker}
\affiliation{Department of Physics, California Institute of Technology, MC 114-36, Pasadena, CA~91125}
\author{Gil Refael}
\affiliation{Department of Physics, California Institute of Technology, MC 114-36, Pasadena, CA~91125}
\author{ Paul M.~Goldbart}
\affiliation{Department of Physics and Institute for Condensed Matter Theory,
University of Illinois at Urbana-Champaign, Urbana, IL~61801, U.S.A.}

\begin{abstract}
Recent experiments on the conductance of thin, narrow superconducting strips have found periodic fluctuations, as a function of the perpendicular magnetic field, with a period corresponding to approximately two flux quanta per strip area~[A. Johansson et al., Phys. Rev. Lett. {\bf 95}, 116805 (2005)].  We argue that the low-energy degrees of freedom responsible for dissipation correspond to vortex motion. Using vortex/charge duality, we show that the superconducting strip behaves as the dual of a quantum dot, with the vortices, magnetic field, and bias current respectively playing the roles of the electrons, gate voltage and source-drain voltage. In the bias-current vs.~magnetic-field plane, the strip conductance displays what we term `Weber blockade' diamonds, with vortex conductance maxima (i.e., electrical resistance maxima) that, at small bias-currents, correspond to the fields at which strip states of $N$ and $N+1$ vortices have equal energy.
\end{abstract}

\maketitle

\noindent%

{\it Introduction -- \/}
It is often effective to characterize strongly correlated quantum systems in terms of the emergent, collective freedoms that describe their low-energy behavior.  Vortices in superconductors constitute the most prominent example of such freedoms, and it has proven useful to address the Kosterlitz-Thouless phase transition exhibited by thin-film superconductors in terms of the statistical mechanics of an interacting plasma of such vortices.  The superfluid-insulator quantum phase transition---exhibited by quite a number of systems~\cite{YazdaniKapitulnik1995, Goldman1999, MasonKapitulnik2001}, particularly granular InO films~\cite{Shahar2007, KovalOvidyahu2008}---is widely suspected to result from a vortex proliferation transition. If so, this behavior would provide a convincing demonstration of vortices behaving quantum-mechanically: not only do they exhibit quantal motion as individuals~\cite{Sahu2009}, but they are also able to Bose-Einstein condense.

Motivated by recent experiments by the Shahar group on superconducting InO strips~\cite{Shahar2005}, which exhibited oscillations in the resistance as a function of magnetic field, in this paper we address such oscillations from the vortex point of view, and show that the notion of a \lq\lq vortex blockade\rq\rq\ allows us to explain such oscillations, in analogy with the Coulomb blockade theory that has been applied widely to electron passage through quantum dots.  Our approach will be applied to the case of a vortex blockade in strips that are narrow.  By using charge-vortex duality, together with energetics arguments, we show that a vortex blockade results in a series of \lq Weber\rq\ diamonds, which are analogs of the Coulomb-blockade diamonds, except for the important distinction that the electrical conductivity is maximal (rather than minimal) inside the Weber diamonds.  We then apply the Beenakker formalism, developed to account for dissipation in quantum-dot transport~\cite{Beenakker}, to the case of a superconducting strip, with the dissipation taking place via the normal modes of the vortex \lq crystal\rq\ in the strip.  Within this dissipative model, we predict that the resistivity of the strip should diverge---possibly observably---at low temperatures, as a power law.  We conclude with a comparison between our results and experimental data, and thus show that our energetics considerations correctly account for period of the magnetoresistance oscillations, observed experimentally.

\begin{figure}[h]
\includegraphics[width=8cm]{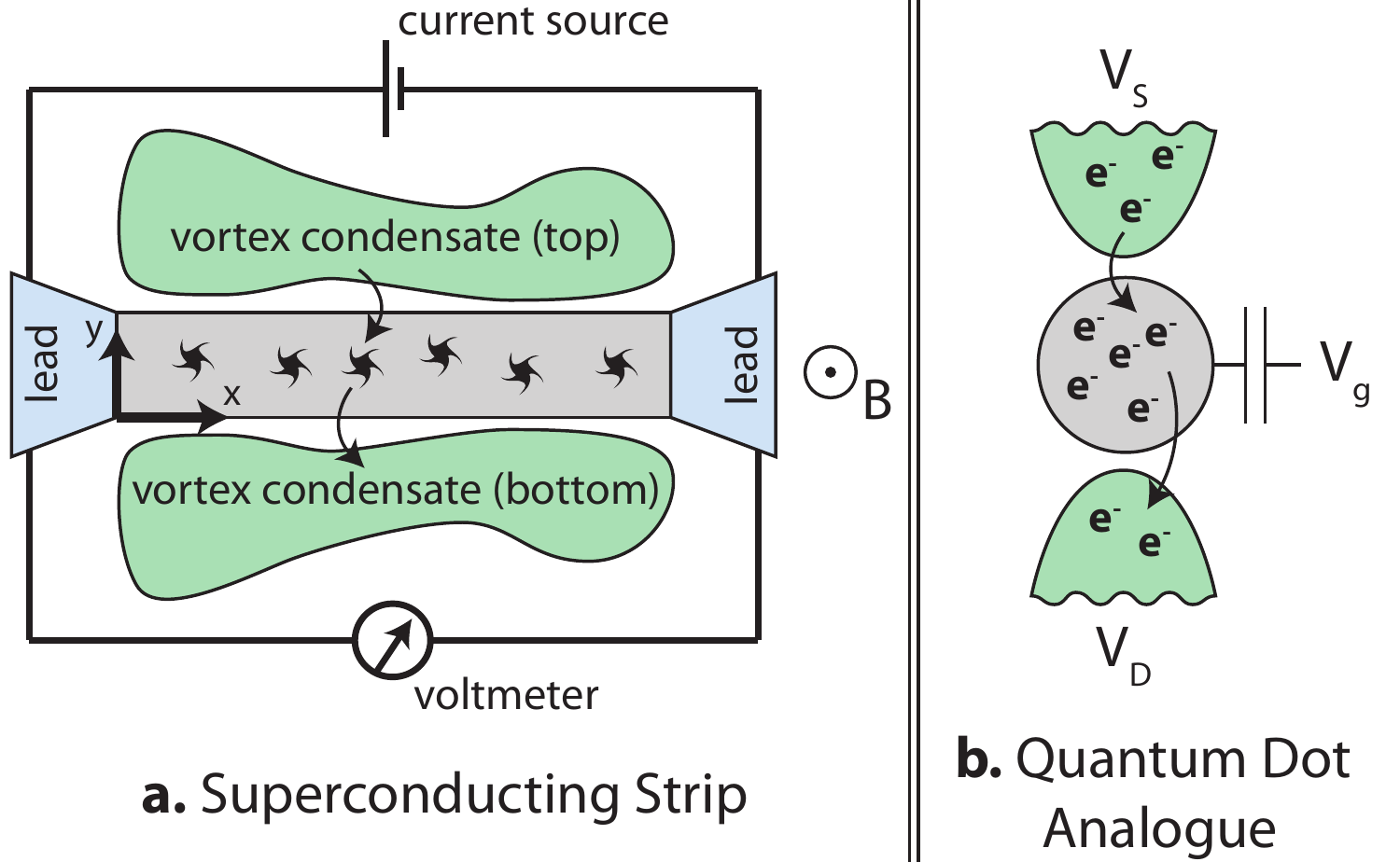}
\caption{(a)~Schematic representation of the experimental setup.  A superconducting strip (gray) is immersed in a magnetic field $B$ pointing out of the plane of the strip.  The strip is contacted via a pair of leads (blue) that are used to pass a bias supercurrent $J$ along the length of the strip.  Simultaneously, the time-averaged voltage is measured using the same leads.  Vortices (depicted by swirls) reside in the strip, but occasionally traverse the strip (cf.~the vortex trajectory indicated by arrows).  Via the Josephson relation, such crossings correspond to voltage spikes, and thus result in dissipation.  (b)~Analogous quantum-dot circuit.  The dot is depicted by the gray disc.  The vortices and the magnetic field in the strip correspond, respectively, to the electrons and the gate voltage $V_g$ of the dot, the latter controlling the mean number of particles on the dot.  The source and drain for the vortices are the vacua adjacent to the strip.  These vacua can be regarded as vortex condensates (green), with the potential energy difference between the condensates set by the bias current.  Thus, $J$ in the vortex analogy corresponds to the source-drain voltage $V_S-V_D$ in the dot (see text for details).}
\label{fig:setup}
\end{figure}

\noindent%
{\it Setup -- \/}
We are concerned with describing the current-voltage characteristics of a superconducting strip that is subject to a perpendicular magnetic field.  The setup that we have in mind is depicted in Fig.~\ref{fig:setup}a.  It consists of a superconducting strip, contacted via a pair of leads that are used to inject a supercurrent into the strip.  The voltage across the strip is read out using the same leads, and we rely on the assumption that the dissipation being probed originates entirely within the superconducting strip and not in the leads.

We make several restrictions on the strip geometry that are pertinent from the vortex perspective.  First, we treat the superconductivity in the strip as being essentially two dimensional, i.e., the strip thickness $t$ is smaller than the coherence length $\xi$.  This restriction implies that vortices in the strip are pancake (i.e., Pearl) vortices, as opposed to line vortices~\cite{Pearl}.  Vortices in thin strips have been investigated previously, using magnetic force imaging~\cite{Martinis}.  These investigations reveal that strips are of two types: narrow and wide.  In narrow strips (i.e., those having a width no greater than several coherence lengths), the vortices always lie in a line along the strip center, like peas in a pod.  For increased magnetic fields, the linear density of the vortices is increased, until an upper critical field is reached.  On the other hand, in wider strips the vortices tend to form Abrikosov-like lattices at higher fields.  In the present work, we concentrate on the case of narrow strips, and use intuition gained from Ref.~\cite{Martinis} to describe the vortices.  We shall address the question of wider strips in a separate article.  Finally, we comment that we consider the strip to be free of sites that would pin vortices, and also to be of finite length (in practice~\cite{Shahar2005}, $\sim 10-100 \, \xi$) so that configurations having distinct numbers of vortices have sufficiently different energies as compared to $k_B T$.

\noindent%
{\it Dual description of a superconducting strip -- \/}
Via charge/vortex duality, we map the behavior of a superconducting strip to that of a quantum dot capable of undergoing a Coulomb blockade (see Fig.~\ref{fig:setup}b).  In this mapping, the superconducting strip plays the role of the quantum dot. The vacua running along either side of the strip act as reservoirs that can exchange vortices with the strip, and thus play the role of the source and drain contacts of a quantum dot. As the vacua do not a have well-defined number of vortices, we describe them as vortex condensates of well-defined phase. The magnetic field controls the chemical potential of the vortices on the strip, and thus acts as the gate voltage of the dot.  Lastly, the bias current that runs along the strip exerts a Magnus force on the vortices, which acts in the direction perpendicular to the current flow.  Effectively, this force creates a potential difference between the vacua on either side of the strip, and thus it acts as the source-drain bias of the dot does.  The transport of vortices across the superconducting strip has a direct consequence for electrical transport within the strip, viz., each time a vortex passes across the strip, the phase between the left and right leads changes by $2 \pi$. The voltage
$V$ between the leads is given by the Josephson relation:
\begin{align}
V=\Phi_0\, \Gamma_N,
\end{align}
where $\Gamma_N$ is the net rate of vortex tunneling, and $\Phi_0\equiv h/2e$ is the superconducting flux quantum.  

Having defined the analogy, we split the problem of computing the vortex conductance (i.e., the electrical resistance) of the strip into two sub-problems.  The first consists of describing the energetics of static vortices.  That is to say, we describe the energy cost of adding an extra vortex to a line of vortices already in the strip.  The second sub-problem consists of utilizing the vortex energetics to construct a master equation describing vortex hopping.  In constructing the master equation we make the typical---and experimentally relevant---assumption that vortex tunneling rates $\Gamma_t$ ($\Gamma_b$) between the strip and the top (bottom) vortex reservoirs are the smallest energy scales in the system, i.e., $\Gamma_t, \Gamma_b \ll k_B T$.  This assumption ensures that after a vortex tunnels onto the strip it is completely de-phased, before it is likely to tunnel again, therefore validating the use of the master equation approach.

\noindent%
{\it Vortex energetics --\/}
To use the aforementioned analogy, we need to quantify the vortex energetics.  To do this, we describe the strip in terms of a phase-only model Hamiltonian,
\begin{align}
H=\frac{\rho}{2} \int dx\,dy\,\left\vert {\bm \nabla} \phi - \frac{2 \pi}{\Phi_0}\, {\bm A} \right\vert^2,
\label{eq:phase}
\end{align}
in which $\rho \sim k_B T_\text{BKT}$ is the superfluid stiffness, $\phi$ is the phase of the order parameter, and ${\bm A}=-B (y-w/2)\hat{\bm e}_x$ is the vector potential in the London gauge.  We use a coordinate system in which $y$ runs from $0$ to $w$ across the width of the strip and the long direction of the strip runs in the $x$ direction, as indicated in Fig.~\ref{fig:setup}a.  We supplement the Hamiltonian with the boundary conditions of no supercurrent through the top or bottom edges of the strip.  The boundary conditions for the left and right ends of the strip depend on the properties of the contacts, and we return to this question, later.  For the purpose of quantitative energetics, we invoke periodic boundary conditions in the long direction.  We note, however, that any choice of boundary conditions that preserves the discreteness of the vortex \lq\lq charging\rq\rq\ energy would result in magnetoresistance oscillations.  To proceed, we break the first sub-problem up into a single-vortex energetics part and an inter-vortex interaction part.

The potential energy of a single vortex in a strip, as a function of its position $y$ across the strip, was thoroughly explored previously~\cite{Likharev}, with a result that comprises four parts, as follows:
(i)~The vortex core energy
$E_\text{core}$,
which is determined via the microscopic theory of the vortex.
(ii)~The energy of interaction of the vortex with its images,
i.e.,
$E_\text{image}=- 2\pi \rho \ln[\sin(\pi \xi/w)/\sin(\pi y/w)]$.
This interaction diverges logarithmically, as the vortex approaches the strip edge, due to the closest image-vortex;  however, the divergence is unphysical, as vortices are finite in size, and one should therefore cut the divergence off at the coherence lengthscale $\xi$. In further calculations, we shall absorb the vortex core energy into the cut-off lengthscale.
(iii)~The part that accounts for the interaction of the vortex with the magnetic field,
i.e.,
$E_{vB}=(2 \pi)^2 (\rho B/\Phi_0)
\left[
 \big(y-w/2\big)^{2}
-\big(\xi-w/2\big)^{2}
\right]$.
(iv)~Finally, the part that describes the potential due to the Magnus force, i.e.,
$E_\text{Magnus}=\Phi_0 (J/w) (y-w/2)$, where $J$ is the bias current.

 \begin{figure}
 \includegraphics[width=8cm]{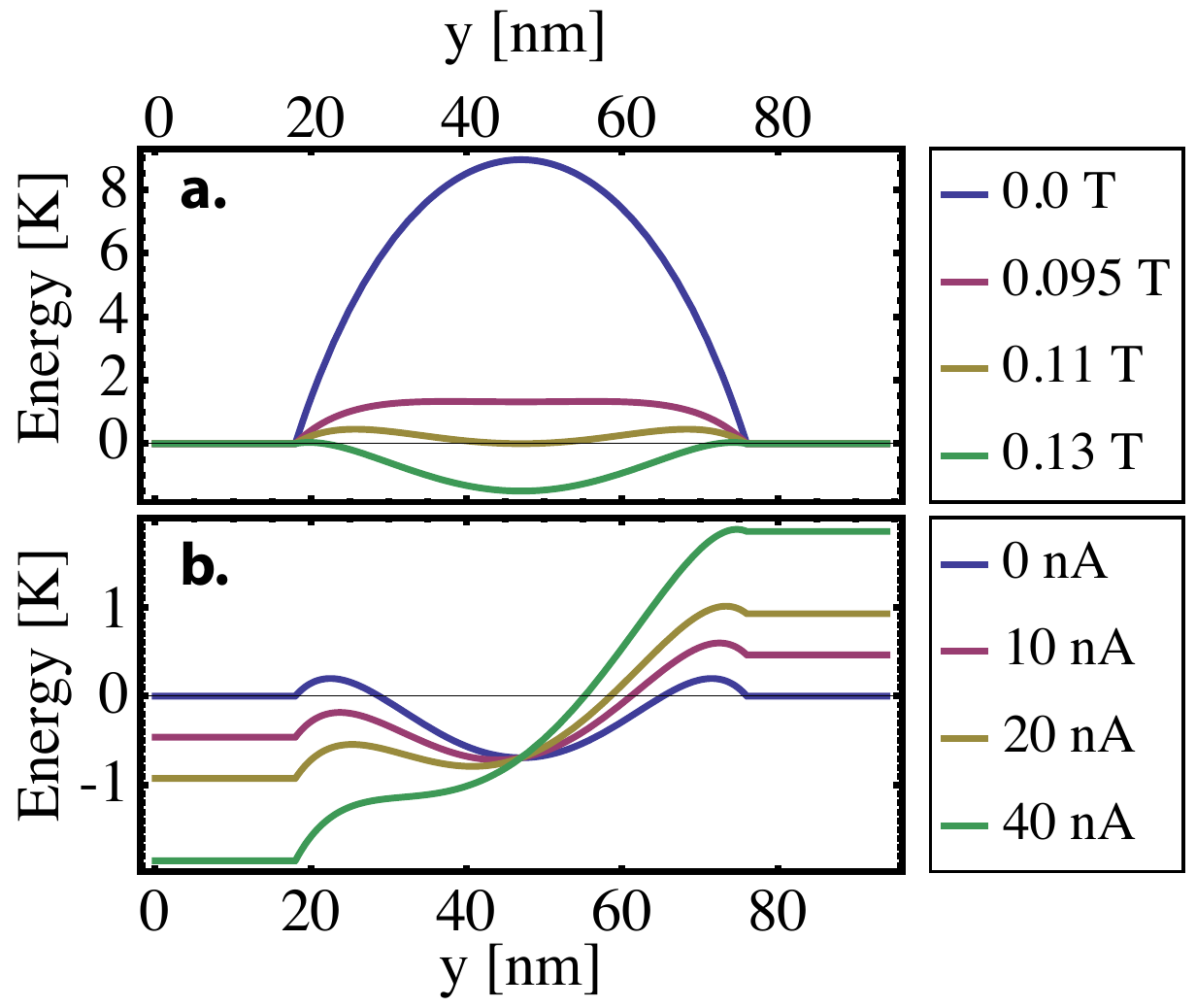}
 \caption{%
 (a)~Vortex potential energy as a function of position across the strip for various magnetic fields for a $94\,\text{nm}$ wide strip with $\xi=18\,\text{nm}$ and $\rho=5\,\text{K}$, (see Ref.~\cite{Likharev}). (b)~Same as~(a), but with nonzero bias current applied along the strip and magnetic field fixed at $0.12\,\text{T}$.
 \label{Fig:Single}}
 \end{figure}
 
The consequences of these four parts are summarized in Fig.~\ref{Fig:Single}a, in which we plot the potential energy of a single vortex in an infinitely long strip, as a function of its position.  For small applied magnetic fields, vortices are unstable everywhere inside the strip.  For larger magnetic fields, the configuration having a vortex trapped in the middle of the strip becomes metastable. At a still larger field (i.e., the lower critical field) the vortex becomes stable, globally.  A bias current along the strip produces a tilting of the vortex potential energy diagram (see Fig.~\ref{Fig:Single}b). In particular, we see that a vortex at the top edge of the strip has a potential energy that differs from that of a vortex at the bottom edge. We identify this energy difference, which corresponds to the work done by the Magnus force as a vortex is moved across the strip, as the difference in potential between the top and bottom vortex reservoirs.

To find the interaction energy between vortices, described by Eq.~\eqref{eq:phase}, we use the conformal mapping
$z=x+iy=\exp({\pi u/w})$
to identify the strip $0<\text{Im}\,u<w$ and the entire upper half-plane
$0<\text{Im}\,z<\infty$.
The mapping works by sending the point at $z=0$ to $u\rightarrow -\infty$, thereby folding the line at
$\text{Im}\,z=0$
in half so as to form the two edges of the strip.
Thus, the boundary condition of no supercurrent through the upper and lower edges of the strip, i.e.,
$\left.\partial_{\text{Im}(u)} \phi(u)\right|_{u=\{0,w\}}$,
becomes the condition of no current through the line
$\text{Im}\,z={0}$, i.e.,
$\left.\partial_{\text{Im}(z)}\phi(z)\right|_{z=0}$.
The phase field
$\phi(z;z_{0})=\text{Im}\left[ \ln(z-z_{0})-\ln(z-{z}_{0}^*)\right]$
of a vortex located at $z_0=x_0 +i y_0$ in the upper half-plane can be found using a single image-vortex.  Next, by using the Cauchy-Riemann equations, we identify the vortex potential $G_\text{half-plane}$ as the real part of the harmonic function whose imaginary part corresponds to the vortex phase field, i.e., 
\begin{align}
G_\text{half-plane}(z;z_{0})\!=\!
2\pi\rho_{s}\text{Re}
\left[
\ln(z-z_{0})\!-\!\ln(z-{z}_{0}^*)
\right].
\end{align}
Applying the conformal mapping, we find the result
\begin{eqnarray}
&&
G_\text{strip}(u;u_{0})=
\\
&&
2\pi\rho_{s}
\text{Re}\,
\big[
\ln(e^{\pi u/w}-e^{\pi u_{0}/w})-
\ln(e^{\pi u/w}-e^{\pi u_{0}^*/w})
\big].
\nonumber
\end{eqnarray}
for the potential $G_\text{strip}$ of a vortex in the strip.
Using this vortex potential, we find that the energy due to inter-vortex interactions is given by
\begin{align}
E_{vv}=\frac{1}{2}
\sum\nolimits_{i\neq j}
G_\text{strip}(u_i;u_j).
\end{align}
We note that the inter-vortex interaction becomes zero for vortices close to the top or bottom edges of the strip. Thus, the vortices in the strip do not influence the potentials of the vortex reservoirs.

\noindent%
{\it Vortex blockade -- \/}
We begin by examining the Weber (cf.~Coulomb) blockade in the absence of a bias current (cf.~source-drain voltage). The key to either blockade is that for generic values of the magnetic field $B$ (cf.~gate voltage $V_g$) the energies for $N$ or $N+1$ vortices (cf.~electrons) to be on the strip (dot) differ, and therefore vortex (cf.~charge) current does not flow. However, for special values of $B$ (cf.~$V_g$), the degeneracy condition
\begin{align}
E_{N}(B,J=0) = E_{N+1}(B,J=0)
\label{eq:period}
\end{align}
is met, allowing vortices (cf.~electrons) to move freely on to and off of the strip (cf.~dot), thus lifting the blockade.

\begin{figure}
\includegraphics[width=7cm]{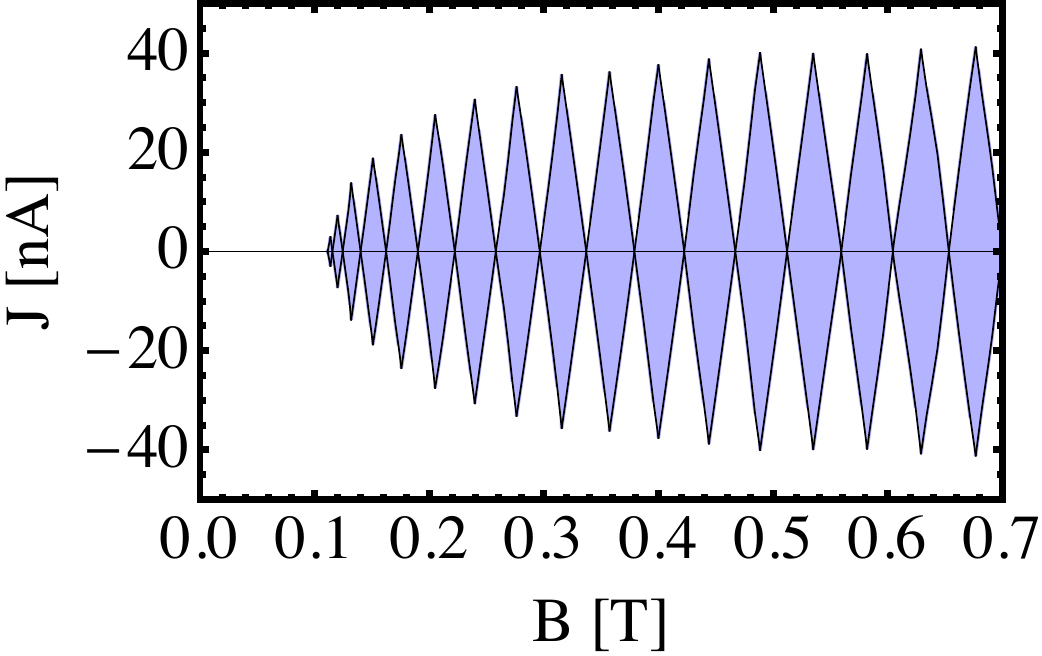}
\caption{Weber diamonds in the bias current vs.~magnetic field plane ($w=94\,\text{nm}$, $L=1\,\mu\text{m}$, $\xi=18\,\text{nm}$, and $\rho=5\,\text{K}$).  Regions inside the diamonds correspond to small electrical resistance along the superconducting strip or, equivalently, to small vortex conductance across it).}
\label{fig:diamonds}
\end{figure}

For the case of nonzero bias current $J$, we must consider separately the processes of adding vortices from the top and bottom reservoirs. These considerations furnish two stability conditions, viz.,
\begin{subequations}
\begin{eqnarray}
E_{N}(B,J) + E_\text{top}    &=& E_{N+1}(B,J),\\
E_{N}(B,J) + E_\text{bottom} &=& E_{N+1}(B,J),
\end{eqnarray}
\end{subequations}
that define the Weber diamonds, in which $E_\text{top}$ and $E_\text{bottom}$ are the potentials of the top and bottom reservoirs. In Fig.~\ref{fig:diamonds} we plot the first few diamonds in the $J$ vs.~$B$ plane.  The regions inside the diamonds correspond to $1, 2,\dots$ vortices on the strip in the blockaded regime (i.e., poor vortex conduction across, and therefore good electrical conduction along, the strip).  
For increased bias currents, the potential difference between the vortex reservoirs overcomes the blockade, driving the strip into the good vortex conduction regime (i.e., to poor electrical conduction along the strip).

In order to estimate the actual magnitude of the conductance of the superconducting strip, we need to extend the static model of the blockade to include the dynamics of vortex hopping and energy dissipation.  For the quantum-dot case, dissipation arises in the leads, which act as \lq\lq decohering baths\rq\rq\ in which the quasiparticles are assumed always to obey the Fermi-Dirac distribution~\cite{Beenakker}.  
In the vortex case, dissipation occurs not in the vortex reservoirs but in the strip itself. 
However, there are many other possible damping mechanisms, such as coupling to the bath of Bogoliubov quasiparticles. The question of which mechanism dominates depends on the microscopic properties of the superconductor.  We shall focus here on the simplest possibility, which is to use as the thermal bath the normal modes of the line of vortices inside the strip.

In our phenomenological approach, we begin by modeling the line of vortices inside the strip as a Luttinger liquid having a single gapless mode, corresponding to the longitudinal sound mode of the one-dimensional vortex crystal. Next, we describe the coupling between the vortex reservoirs and the Luttinger liquid by introducing a model Hamiltonian that accounts for vortex tunneling between them:
\begin{align}
H_\text{tun}= \int dx \, b(x) \left(t_t\,e^{i \theta_t} + t_b\,e^{i \theta_b} \right) + {\textrm h.c.},
\end{align}
where $b(x)$ is the operator that annihilates a vortex at position $x$ along the strip.  Following Beenakker's approach to quantum dots~\cite{Beenakker}, which invokes Fermi's Golden Rule, we obtain a master equation for the probabilities $P_0, P_1,\dots$ of having $0,1,\dots$ vortices on the strip: 
\begin{widetext}
\begin{subequations}
\begin{align}
\partial_t P_N =&  \phantom{+}  \sum_{\epsilon_i, \epsilon_f} \sum_{\sigma=\pm 1} 
P_{N+\sigma} \, e^{-\beta \epsilon_i}
\left\vert_{N}\langle \epsilon_f | H_\text{tun} | \epsilon_i \rangle_{N+\sigma} \right\vert^2 \sum_{\alpha \in \{\text{t},\text{b}\}}\Gamma_\alpha \, \delta\left[\left(E_{N+\sigma}+\epsilon_i\right)-\left(E_N+\epsilon_f\right)-\sigma \, {\cal E}_\alpha\right] \\
&- \sum_{\epsilon_i, \epsilon_f} \sum_{\sigma=\pm 1} 
P_{N} \, e^{-\beta \epsilon_i}
\left\vert_{N+\sigma}\langle \epsilon_f | H_\text{tun} | \epsilon_i \rangle_{N} \right\vert^2 \sum_{\alpha \in \{\text{t},\text{b}\}}\Gamma_\alpha \, \delta\left[\left(E_{N}+\epsilon_i\right)-\left(E_{N+\sigma}+\epsilon_f\right)-\sigma \, {\cal E}_\alpha\right]
\end{align}
\end{subequations}
\end{widetext}
where $\left\vert \epsilon_i \right\rangle_N$ and $\left\vert \epsilon_f \right\rangle_N$ correspond to the states of the Luttinger liquid (with $N$ vortices), ${\cal E}_t$ and ${\cal E}_b$ are the potentials of the top and bottom vortex reservoirs, and the $\delta$-functions account for energy conservation. 

In the regime of small bias-supercurrent $J$ (cf.~low source-drain bias voltage),
we can solve the master equation and use the solution to recover the vortex conductivity $\sigma$ (cf.~electrical resistivity): 
\begin{align}
\sigma(J) \sim \frac{\Gamma_t\,\Gamma_b}{\Gamma_t + \Gamma_b} 
\int dt \, e^{i J t} \int \frac{dx\, dx'}{\left(t^2+(x-x')^2\right)^{1/4K}},
\label{VB}
\end{align}
where $K$ is the corresponding Luttinger parameter. Here, we have used the standard trick~\cite{Nazarov} of converting the $\delta$-functions into time integrals to recover the retarded Green functions.  In general, the $(x\,x)'$ integrals will be cut off by the system size at large argument and by the inter-vortex spacing at small argument.  Finally, we comment that the boundary conditions for the Luttinger liquid are determined by the coupling of the vortices in the strip to the degrees of freedom in the leads. 

\begin{table}
\begin{tabular}{|c|c|c|c|c|c|}
\hline
sample & $w$ [nm] & $L$ [$\mu$m] & $P$ [T] & $P_\text{geometric}$ [T] & $\xi$ [nm]\\
\hline
1 & 40 & 1.5 & 0.083 & 0.034 & 10 \\
2 & 94 & 1.0 & 0.048 & 0.022 & 17 \\
3 & 94 & 1.0 & 0.046 & 0.022 & 16 \\
4 & 94 & 1.0 & 0.043 & 0.022 & 15 \\
\hline
\end{tabular}
\caption{Coherence length $\xi$ (a fitting parameter) required to reproduce the experimentally measured period $P$ using our theoretical model for four samples~\cite{Sample5} of widths $w$ and lengths $L$ (Table~I of Ref.~\cite{Shahar2005}). In each case, we fit the experimental period at the intermediate value of magnetic field $B \approx 1.5\,\text{T}$.  For comparison, we also provide the geometric period $P_\text{geometric}=\left(\Phi_0\,w\,L\right)^{-1}$. For each sample, we find a reasonable value of coherence length, which should lie in the range from $10\,\text{nm}$ to $30\,\text{nm}$, according to Ref.~\cite{Shahar2005}. }
\label{table:exp}
\end{table}

\noindent%
{\it Comparison with experiment -- \/} Using Eq.~\eqref{eq:period}, we can estimate theoretically the period of the magnetoresistance oscillations for the sample geometries of Ref.~\cite{Shahar2005}.  The only undetermined parameter in the theoretical estimate is the coherence length. Treating the coherence length as a fitting parameter, we were able to fit the periods of all four narrow-strip samples (Table~\ref{table:exp}).  We remark that the experimentally observed period is always longer than the geometric period associated with adding a flux quantum to the area of the strip (Table~\ref{table:exp}). Our theory implies that the effective width of the strip is narrower than its geometrical width, due to the strong attraction of vortices to the strip edges by their images (i.e., $E_\text{image}$), thus resolving the period puzzle. The experimental signal shows only mild magnetoresistance oscillations,  rather than the pronounced vortex-blockade peaks that we predict.  We suspect that this smearing is a result of $\Gamma_t$ and $\Gamma_b$ being large, and thus the strip operating in between the open- and closed-dot regimes. Alternatively, the smearing may be a result of thermal broadening. If so, by lowering the temperature the blockade features are expected to be strengthened.

\noindent%
{\it Concluding remarks -- \/} We have demonstrated that magnetoresistance oscillations in superconducting strips can be readily be described using vortex---as opposed to Cooper-pair or charge---coordinates. Via an analogy with the physics of quantum dots, we have constructed a model for the \lq Weber blockade\rq\ of a superconducting strip that captures the essential features of the magnetoresistance oscillations observed experimentally. The
exact dissipation mechanism remains to be understood; it will
determine the shape of the Coloumb blockade peaks predicted in
Eq. (\ref{VB}) but not their locations. Alternatively, one could use
the experimental blocakde signatures in order to investigate the
nature of dissipation, according to Eq. (\ref{VB}).

\noindent%
{\it Acknowledgments -- \/} 
The authors thank D.~Shahar and J.~Meyer for encouragement and useful discussions.
This work was supported by DOE~DE-FG02-07ER46453 (PMG), the Research
Corporation, Packard Foundation and the Sloan Foundation (GR), and the
Sherman-Fairchild foundation (DP).  
The authors thank for its hospitality the Aspen Center for Physics, 
where part of this work was carried out.
During the preparation of this manuscript, we became aware of work by Y.~Atzmon and E.~Shimshoni that approaches the issue of the magnetoresistance oscillations from the point of view of the superconductor-insulator transition~\cite{Shimshoni}. This approach and ours lead to similar physical conclusions.

\end{document}